\documentclass[12pt,preprint]{aastex}

\slugcomment{2012, ApJ, 745, L22, doi:10.1088/2041-8205/745/2/L22}

\begin{document}


\title{Observation of TeV gamma rays from  the Cygnus region with the ARGO-YBJ experiment}


\author{B.~Bartoli\altaffilmark{1,2},
 P.~Bernardini\altaffilmark{3,4},
 X.J.~Bi\altaffilmark{5},
 C.~Bleve\altaffilmark{3,4},
 I.~Bolognino\altaffilmark{6,7},
 P.~Branchini\altaffilmark{8},
 A.~Budano\altaffilmark{8},
 A.K.~Calabrese Melcarne\altaffilmark{9},
 P.~Camarri\altaffilmark{10,11},
 Z.~Cao\altaffilmark{5}
% ...
\\ (The ARGO-YBJ Collaboration)}


 \altaffiltext{1}{Dipartimento di Fisica dell'Universit\`a di Napoli
                  ``Federico II'', Complesso Universitario di Monte
                  Sant'Angelo, via Cinthia, 80126 Napoli, Italy.}
 \altaffiltext{2}{Istituto Nazionale di Fisica Nucleare, Sezione di
                  Napoli, Complesso Universitario di Monte
                  Sant'Angelo, via Cinthia, 80126 Napoli, Italy.}
 \altaffiltext{3}{Dipartimento di Fisica dell'Universit\`a del Salento,
                  via per Arnesano, 73100 Lecce, Italy.}
 \altaffiltext{4}{Istituto Nazionale di Fisica Nucleare, Sezione di
                  Lecce, via per Arnesano, 73100 Lecce, Italy.}
 \altaffiltext{5}{Key Laboratory of Particle Astrophysics, Institute
                  of High Energy Physics, Chinese Academy of Sciences,
                  P.O. Box 918, 100049 Beijing, P.R. China.}
 \altaffiltext{6}{Dipartimento di Fisica Nucleare e Teorica
                  dell'Universit\`a di Pavia, via Bassi 6,
                  27100 Pavia, Italy.}
 \altaffiltext{7}{Istituto Nazionale di Fisica Nucleare, Sezione di Pavia,
                  via Bassi 6, 27100 Pavia, Italy.}
 \altaffiltext{8}{Istituto Nazionale di Fisica Nucleare, Sezione di
                  Roma Tre, via della Vasca Navale 84, 00146 Roma, Italy.}
 \altaffiltext{9}{Istituto Nazionale di Fisica Nucleare - CNAF, Viale
                  Berti-Pichat 6/2, 40127 Bologna, Italy.}
 \altaffiltext{10}{Dipartimento di Fisica dell'Universit\`a di Roma ``Tor Vergata'',
                   via della Ricerca Scientifica 1, 00133 Roma, Italy.}
 \altaffiltext{11}{Istituto Nazionale di Fisica Nucleare, Sezione di
                   Roma Tor Vergata, via della Ricerca Scientifica 1,
                   00133 Roma, Italy.}
% ...

\begin{abstract}
We report the observation of TeV $\gamma$-rays from the Cygnus region using the ARGO-YBJ data collected from 2007 November  to 2011 August.
Several TeV sources are located in this region including the two bright extended MGRO J2019+37 and MGRO J2031+41.
According to the Milagro data set, at 20 TeV MGRO J2019+37 is the most significant source apart from the Crab Nebula.
No signal from MGRO J2019+37 is detected by the ARGO-YBJ experiment, and the derived flux upper limits at 90\% confidence
level for all the events above 600 GeV with medium energy of 3 TeV are lower than the Milagro flux,
implying that the source might be variable and hard to be identified as a pulsar wind nebula. The only statistically significant (6.4 standard deviations) $\gamma$-ray
signal is found from MGRO J2031+41, with a flux consistent with the
measurement by Milagro.
\end{abstract}


\section{Introduction}
The Cygnus region  is  the brightest diffuse
$\gamma$-ray emitting region in the northern sky as
observed by both $Fermi$ \citep{abdo11} and EGRET \citep{hunter97}.
Complex features have been observed in the wavelength bands of radio,
infrared, X-rays, and $\gamma$-rays. This region is rich in potential cosmic-ray acceleration sites,
e.g., Wolf-Rayet stars, OB associations and supernova remnants. Recently, 24 $\gamma$-ray sources,
including 7 pulsars,  have been detected using $Fermi$ Large Area Telescope (LAT) two-year data within the region with
$65^{\circ}<l<85^{\circ}$ and $-3^{\circ}<b<3^{\circ}$ \citep{abdo11}. These
are considered candidate sources of very high energy (VHE) $\gamma$-rays.
The Cygnus region is,
therefore, a natural laboratory to study the origin of cosmic rays.

Several VHE $\gamma$-ray  sources have been detected within the Cygnus region
in the past decade. The first was TeV J2032+4130, discovered by the HEGRA
collaboration \citep{aharon02,aharon05}
and confirmed by the experiments Whipple \citep{konop07} and
MAGIC \citep{albert08}. Its extension  is estimated to be about 0.1$^{\circ}$.
The power-law spectral index is about $-$2.0 and the integral flux above 1 TeV
is 3$-$5\% that of the Crab flux.
MGRO J2031+41, detected by the Milagro experiment at 20 TeV \citep{abdo07a},
is spatially consistent with the source TeV J2032+4130, while the measured
extension is much larger with a diameter of 3.0$^{\circ}$$\pm$0.9$^{\circ}$.
This source is likely to be associated with the pulsar 2FGL J2032.2+4126,
detected by $Fermi$ \citep{abdo11}.

Evidence of TeV emission at the 4.0 standard deviation (s.d.) level after an X-ray flare  from Cyg X-1 was
observed by the MAGIC experiment  on 2006 September 24  \citep{albert06}.

The source VER J2019+407 was discovered in a survey of the Cygnus region by
the VERITAS experiment \citep{wein09}. The measured  extensions are
$0.16^{\circ}\pm0.028^{\circ}$ and
$0.11^{\circ}\pm0.027^{\circ}$ along the major and minor axes, respectively.
This source is spatially coincident with the $Fermi$ source 2FGL J2019.1+4040, which is  potentially associated with a supernova remnant or a
pulsar wind nebula (PWN) \citep{abdo11}.

During the deep VERITAS
observations of the Cyg OB1 region, a point source VER J2016+372 was
discovered at the location of CTB 87 \citep{aliu11}.
The flux is about 1\% that of the Crab and the spectral index is about $-$2.1 from a preliminary analysis.

This region also contains the bright unidentified source MGRO J2019+37,
which was detected by the Milagro experiment at 20 TeV \citep{abdo07a}
and is the most significant source in the Milagro data set
apart from the Crab Nebula. Its extension is
$\sigma=0.32^{\circ}$$\pm$0.12$^{\circ}$  in a symmetric two-dimensional Gaussian
\citep{abdo07b}, which has 68\% of the events  contained in a region with an angular diameter of
1.1$^{\circ}$$\pm$0.5$^{\circ}$ \citep{abdo07a}.
The spectrum of this source is hard with an index of $-$1.83 and an
exponential cutoff at 22.4 TeV \citep{smith09}.
At the location of MGRO J2019+37, a 2.2 s.d. signal corresponding to
30\% Crab unit was observed by the Tibet AS$\gamma$ experiment \citep{ameno10}.
However, about 0.9$^{\circ}$ away, a possible source was detected
\citep{wang07}. The  MGRO J2019+37 is spatially coincident with the  $Fermi$ source  2FGL J2018.0+3626 and pulsar
2FGL J2021.0+3651 \citep{abdo11}.
VERITAS has surveyed this region, but no emission from MGRO J2019+37 has been
detected \citep{wein09}. Recently, an in-depth observation of the Cyg OB1
region has been carried out by VERITAS, which unveiled complex TeV emission
around MGRO J2019+37 \citep{aliu11}.

Among the four known VHE $\gamma$-ray sources inside the Cygnus region,
MGRO J2019+37 is enigmatic due to its high flux not confirmed by other VHE
$\gamma$-ray detectors. The measurement of the energy spectrum or an upper
limit around several TeV is therefore very useful to understand the nature of the source and its emission
mechanism. The ARGO-YBJ experiment is an air shower array with large
field of view and can continuously monitor the northern sky. The total
exposure to the Crab Nebula reaches about 1200 days and its photon flux has been detected with a statistical
significance of 17  s.d. at energies around 1 TeV,
which is comparable with the eight-year value of 17.2 s.d. obtained at energies around 35 TeV by
Milagro \citep{abdo09}.
This work presents the observation
results for the Cygnus region, including sources MGRO J2031+41 and
MGRO J2019+37, with the ARGO-YBJ experiment.

\section{The ARGO-YBJ experiment}
The ARGO-YBJ experiment, located in Tibet, China, at an altitude of 4300 m a.s.l., is the result of a collaboration among Chinese and Italian institutions and is designed for VHE $\gamma$-ray astronomy and cosmic ray observations. The detector consists of a single layer of Resistive Plate Chambers (RPCs), which
are equipped with pick-up
strips (6.75 cm $\times$ 61.80 cm  each). The logical OR of the signal from eight neighboring strips
constitutes a logical pixel (called a  ``pad'') for triggering and timing
purposes.
One hundred thirty clusters (each composed of 12 RPCs) are installed to form a carpet
of about 5600 m$^{2}$ with an active area of $\sim$93\%. This central
carpet is surrounded by 23 additional clusters (``guard ring'') to
improve the reconstruction of the shower core location. The total area of the
array is  110 m $\times$ 100 m. More details about the detector and
the RPC performance can be found in \citep{aielli06,aielli09b}.

The ARGO-YBJ detector is operated by requiring the number of fired pads ($N_{pad}$)
at least 20   within 420 ns on the entire carpet detector.
The high granularity of the apparatus permits a detailed space-time reconstruction of the
shower profile and therefore of the incident direction of the primary particle. The arrival
time of the particles is measured  with a resolution of about
1.8 ns \citep{aielli09b}. In order to calibrate the 18,360 TDC
channels, we have developed a method using  cosmic ray showers \citep{he07}.
The calibration precision is 0.4 ns and the procedure is applied every month \citep{aielli09}.

The central 130 clusters began taking data in 2006 July, and the ``guard ring'' was merged
into the DAQ stream in 2007 November. The trigger rate is 3.5 kHz with a dead time of 4\%
and the average duty cycle is higher than $86\%$. The angular resolution, pointing accuracy and stability of
the ARGO-YBJ detector array have been thoroughly tested by
measuring the shadow of the Moon in cosmic rays \citep{barto11b}.
The point-spread function (PSF) is quantified using a parameter $\psi_{70}$  as the opening angle containing 71.5\% of the events.
For $N_{pad}>1000$, $\psi_{70}$ is
0.47$^{\circ}$, while at $N_{pad}\sim 20$ $\psi_{70}$ becomes 2.8$^{\circ}$ \citep{barto11,barto11b}.
This measured angular resolution refers to cosmic ray induced
air showers, and it is smaller by 30$-$40\%  for $\gamma$-rays.

\section{Data analysis}
The ARGO-YBJ data used in this analysis were collected from 2007 November  to 2011 August.
The total effective observation time is 1182.0 days. To achieve a good angular resolution,
the event selections used in \citep{barto11} are applied here and only events with zenith angle
less than 50$^{\circ}$ are used.   The total number of events after filtering used in this
work is 1.97$\times$10$^{11}$. The opening angle $\psi_{70}$ for events with $N_{pad}>60$ is 1.36$^{\circ}$.
In order to obtain a sky map, an area centered at the source
location in celestial coordinates (right ascension and declination) is divided into a grid of
$0.1^{\circ}\times0.1^{\circ}$ bins and filled with detected events according to their
reconstructed arrival directions. To extract  the excess of $\gamma$-rays from each bin, the
``direct integral method'' \citep{fleysher04} is adopted in order to estimate the number
of cosmic ray background events in the bin. To remove the effect of cosmic ray anisotropy on a spatial scale of $11^{\circ}\times11^{\circ}$,
a correction procedure described in \citep{barto11} has been applied. To take into
account the PSF of the ARGO-YBJ detector, the events in a circular area centered on the bin
with an angular radius of 1.3$\psi_{70}$, are summed together with a weight of the Gaussian-shaped PSF.
Equation (17) in \citep{li83} is used to estimate the
significance of the excess in each bin.

With this data analysis, the significance of the excess observed from the direction of  the Crab Nebula is 17 s.d., which indicates that the cumulative 5 s.d. sensitivity of ARGO-YBJ has reached 0.3 Crab unit for point sources \citep{chensz11}.  For an extended source with a symmetric two-dimensional Gaussian shape with  $\sigma=0.32^{\circ}$, the sensitivity is degraded by a factor of 11\%.

\section{Results and discussion}
The significance map of the Cygnus region as observed by ARGO-YBJ using events with $N_{pad}>20$
is shown in Figure 1. For
comparison, the 4 known TeV sources and 24 GeV sources in  the second $Fermi$ LAT catalog  are
marked in the figure. An excess is observed over a large part of
the Cygnus region, which indicates a possible diffuse $\gamma$-ray
emission. An analysis of the diffuse $\gamma$-ray emission
using ARGO-YBJ data can be found in \citep{ma11}. The highest
significance value is 6.4 s.d. at
(307.85$^{\circ}$, 41.75$^{\circ}$), consistent with the position
of VHE sources MGRO J2031+41 and TeV J2032+4130.
No evidence of an emission above 3 s.d. is found
at the location of MGRO J2019+37.

\subsection{MGRO J2031+41}
The intrinsic extension of MGRO J2031+41 is determined by fitting the distribution of $\theta^{2}$ for the events exceeding the background as shown in Figure 2, where $\theta$ is the angular distance of each event to the position of TeV J2032+4130. Only events with $N_{pad}>60$ are used in this fit, where $N_{pad}$ is the number of fired pads. To fit the data, a set of $\gamma$-rays is generated taking into account the spectral energy distribution (SED), the intrinsic source extension, and the detector PSF. The extension is estimated, by minimizing the $\chi^2$ between the data and the generated events, between 0$^{\circ}$ and 1$^{\circ}$ with steps of 0.1$^{\circ}$.
Assuming the background spectral index $-$2.8, the intrinsic extension is determined to be $\sigma_{ext}=(0.2_{-0.2}^{+0.4})^{\circ}$. It is found that the dependence on the SED is negligible within the uncertainties. This result is consistent with the estimation by the MAGIC and HEGRA experiments,
i.e., (0.083$\pm$0.030)$^{\circ}$ and (0.103$\pm$0.025)$^{\circ}$, respectively.

Assuming an intrinsic extension $\sigma_{ext}=0.1^{\circ}$, we estimate the spectrum of MGRO J2031+41 using the ARGO-YBJ data by a conventional fitting method described in \citep{barto11}. In this procedure, the expectation function is generated by sampling events in the energy range from 10 GeV to 100 TeV and taking into account the detailed ARGO-YBJ detector response to the events assuming a power law with its spectral index as a parameter. We define four intervals with $N_{pad}$ of 60$-$99, 100$-$199, 200$-$499, and $\geq$500.
The best fit to the SED is shown in Figure 3. The differential flux (TeV$^{-1}$ cm$^{-2}$ s$^{-1}$) in the energy range from 0.6 TeV to 7 TeV is
\begin{equation}
\frac{dN}{dE dA dt}=(1.40\pm0.34)\times10^{-11}(\frac{E}{1TeV})^{-2.83\pm0.37}.
\end{equation}
The integral flux is 31\% that of the Crab at energies above 1 TeV, which is higher than the flux of
TeV J2032+4130 as determined by the HEGRA and MAGIC experiments, i.e. 5\% and 3\% Crab unit, respectively.
However, this measurement is   in agreement with the Milagro new result \citep{boname11}, also shown in Figure 3.

The reason for the discrepancy between the fluxes measured by Cherenkov telescopes and extensive air shower arrays is still unclear.
A  contribution is expected from the diffuse $\gamma$-ray flux produced by cosmic rays interacting with matter  in the Galaxy plane.
According to the measurement  of diffuse $\gamma$-ray flux from the Cygnus region using ARGO-YBJ data \citep{ma11}, this contribution
to the measured flux from MGRO J2031+41 at energy above 1 TeV is about 10\%. Moreover,
an estimate of the systematic error is described in \citep{aielli10}. With an incomplete list of the sources of systematics,
such as the time resolution variation, event rate variation with environment parameters, and pointing error, the error is found to be less than 30\%.
Due to the limited angular resolution of the ARGO-YBJ detector, nearby sources could contribute to the flux. For TeV J2032+4130, this contribution must be very small, because there is no source in the $3^{\circ}\times3^{\circ}$ field of view as seen by  HEGRA and MAGIC except TeV J2032+4130 itself \citep{aharon05,albert08}.
Thus, the contribution from diffuse $\gamma$-ray emission, nearby sources, and systematic uncertainty are not enough to explain the discrepancy.

\subsection{MGRO J2019+37}
No excess above 3 s.d. is detected inside the Cyg OB1 region,
even considering extended sources.
Taking into account the position uncertainty reported by Milagro, the bin with the maximum significance within 0.3$^{\circ}$ from MGRO J2019+37 is used to estimate the upper limits.
The flux upper limits at 90\% confidence level (c.l.) are shown in Figure 4 assuming the SED  reported in \citep{smith09} and
the extension  $\sigma=0.32^{\circ}$ given in  \citep{abdo07b}, respectively.
At energies above 5 TeV, the ARGO-YBJ exposure is still   insufficient to reach a firm conclusion. As regards the emission   at lower energies, the ARGO-YBJ observation does not confirm the spectrum determined by Milagro \citep{smith09}. Taking into account that the two observations differ in time by several years, this discrepancy might indicate a variation in the $\gamma$-ray flux of the source.

The VERITAS experiment carried out a fine scanning of the Cygnus region. With a sensitivity of about 10\% Crab unit,
there was no significant signal found in the direction of MGRO J2019+37 \citep{wein09}.
With a deeper survey corresponding to a sensitivity of about 1\% Crab unit, some faint sources were found in this region \citep{aliu11}.
The estimated flux is much weaker than that determined by
the Milagro experiment.

Considering the source extension $\sigma=0.32^{\circ}\pm0.12^{\circ}$ and the distance of the Cygnus region 1$-$2 kpc, the source radius is estimated to be 4$-$15 pc, implying that the variation timescale should be longer than 13$-$49 years. The observation by the ARGO-YBJ experiment is about five years later than that by Milagro. A flux variation over the whole extended region cannot be completely excluded. If the flux variation were dominated by a smaller region in the source area, the picture could be more reasonable.
In such a scenario, however, identifying  MGRO J2019+37 as a PWN could  be a dilemma  because it otherwise should have a  steady flux.

\section{Conclusions}
Since 2007 November   the ARGO-YBJ experiment is monitoring with high duty cycle the northern sky
at TeV photon energies.
Using data up to 2011 August, we have observed the Cygnus region,
inside which two bright VHE
extended $\gamma$-ray sources have been detected by the Milagro experiment.
An excess with statistical significance of 6.4 s.d. is detected from the direction of MGRO J2031+41, consistent with the Milagro observation, but with a flux higher than that measured by HEGRA and MAGIC. The source location and extension are, however, consistent with those of TeV J2032+4130. It is not easy to assess
the origin of this discrepancy.
No signal from MGRO J2019+37 is detected, and the derived upper limits at 90\% c.l. are lower than the Milagro flux at energies below 5 TeV.
This result could be explained invoking a source variability, making difficult in that case to identify the source as a pulsar wind nebula.
In conclusion, further observations and attention to the Cygnus region are needed since it is found to be complex in the VHE domain.

\acknowledgments
 This work is supported in China by NSFC (No.10120130794),
the Chinese Ministry of Science and Technology, the
Chinese Academy of Sciences, the Key Laboratory of Particle
Astrophysics, CAS, and in Italy by the Istituto Nazionale di Fisica
Nucleare (INFN).

We also acknowledge the essential supports of W.Y. Chen, G. Yang,
X.F. Yuan, C.Y. Zhao, R. Assiro, B. Biondo, S. Bricola, F. Budano,
A. Corvaglia, B. D'Aquino, R. Esposito, A. Innocente, A. Mangano,
E. Pastori, C. Pinto, E. Reali, F. Taurino and A. Zerbini, in the
installation, debugging and maintenance of the detector.


\clearpage
\begin{figure}
\plotone{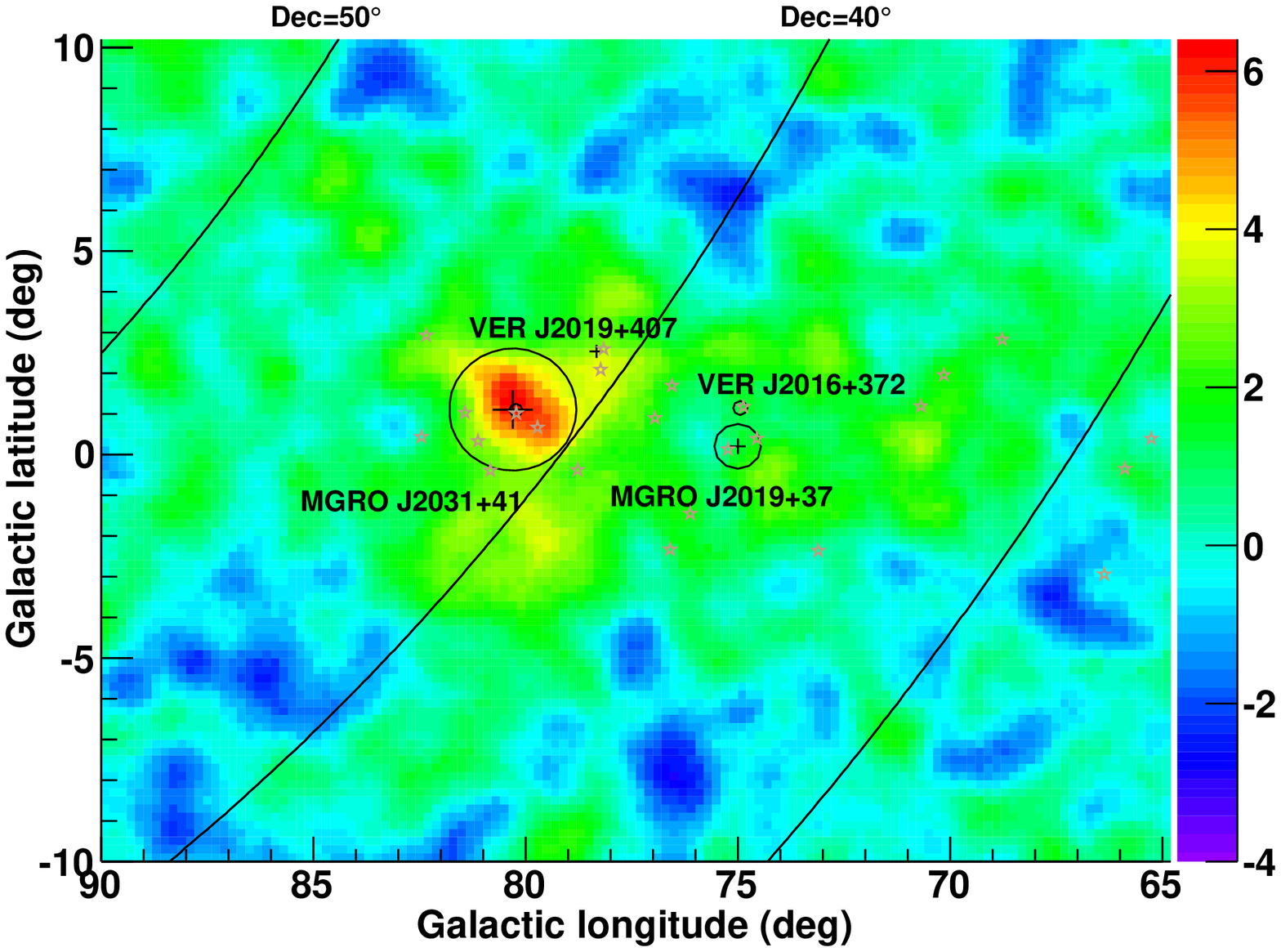}
\caption{Significance map of the Cygnus Region as observed by the ARGO-YBJ
experiment. The four known VHE $\gamma$-ray source are reported.
  The errors on the MGRO source positions are marked with crosses, while the circles indicate their intrinsic sizes \citep{abdo07a,abdo07b}. The cross for VER J2019+407 indicates its extension \citep{wein09}. The source  VER J2016+372 is marked with small circles without position errors. The small circle within the errors of MGROJ2031+41 indicates position and extension of the source TeV J2032+4130 as estimated by the MAGIC collaboration \citep{albert08}. The open stars mark the location of the 24 GeV sources in the second $Fermi$ LAT catalog.}
\label{fig1}
\end{figure}

\begin{figure}
\plotone{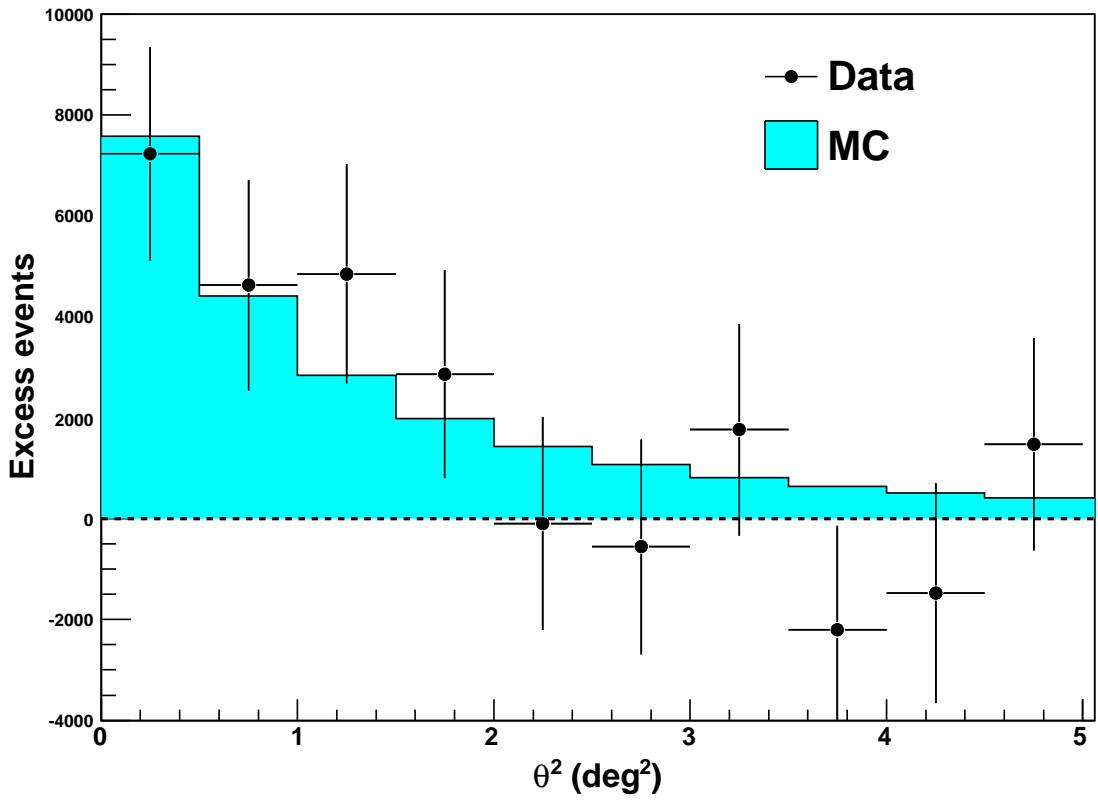}
\caption{Distribution of $\theta^{2}$ for the number of excess events around
TeV J2032+4130. The filled region is the best fit to simulated data.}
\label{fig2}
\end{figure}

\begin{figure}
\plotone{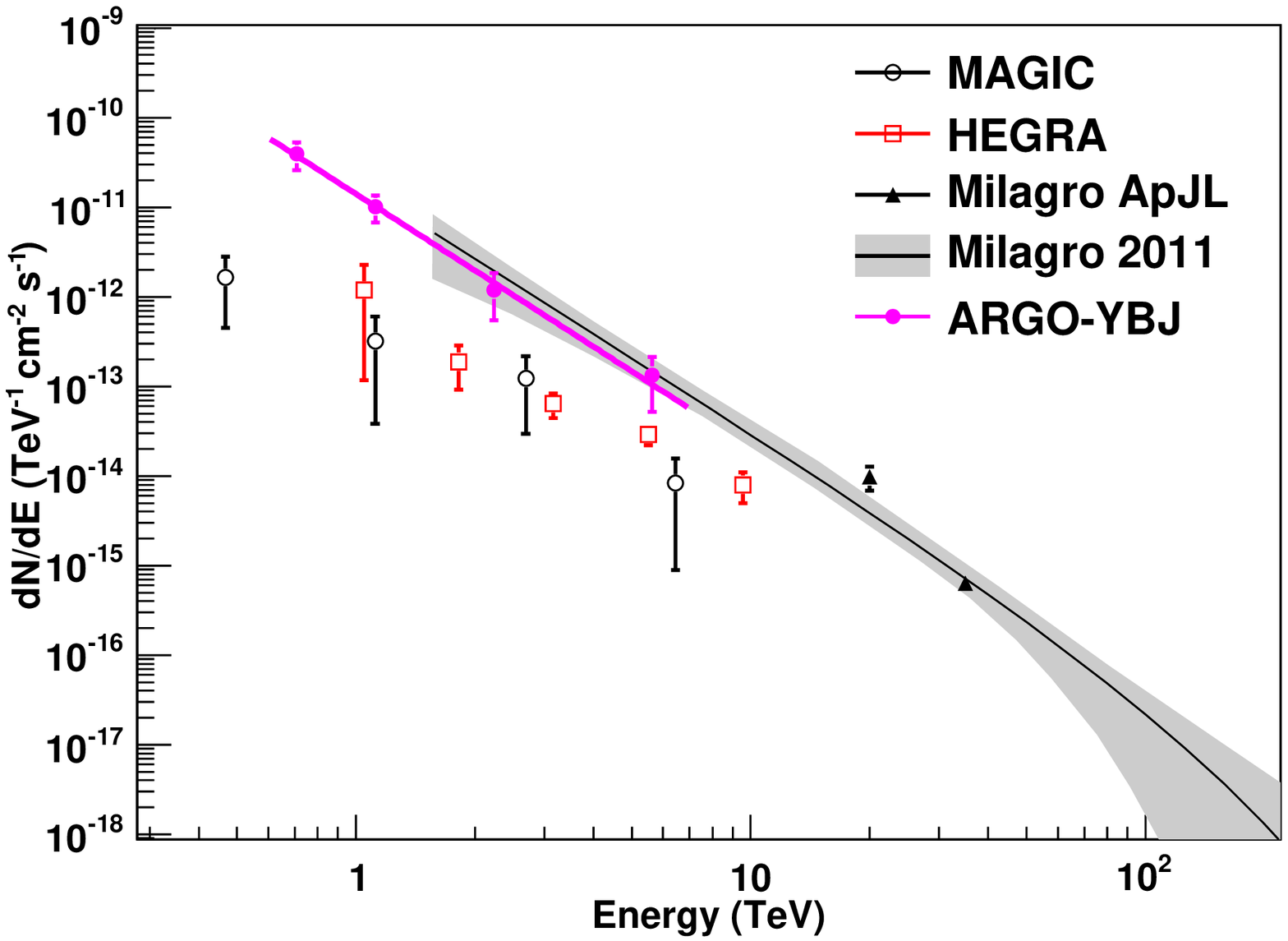}
\caption{Energy density spectrum from TeV J2032+4130/MGRO J2031+41 as
measured by the ARGO-YBJ experiment (magenta solid line). The spectral
measurements of HEGRA \citep{aharon05} and MAGIC \citep{albert08} are also reported for comparison. The solid line and shaded area indicate the differential energy spectrum and the 1 s.d. error region as recently determined
by the Milagro experiment \citep{boname11}. The two triangles give the previous
flux measurements by Milagro at 20 TeV \citep{abdo07a} and
35 TeV \citep{abdo09}.}
\label{fig3}
\end{figure}

\begin{figure}
\plotone{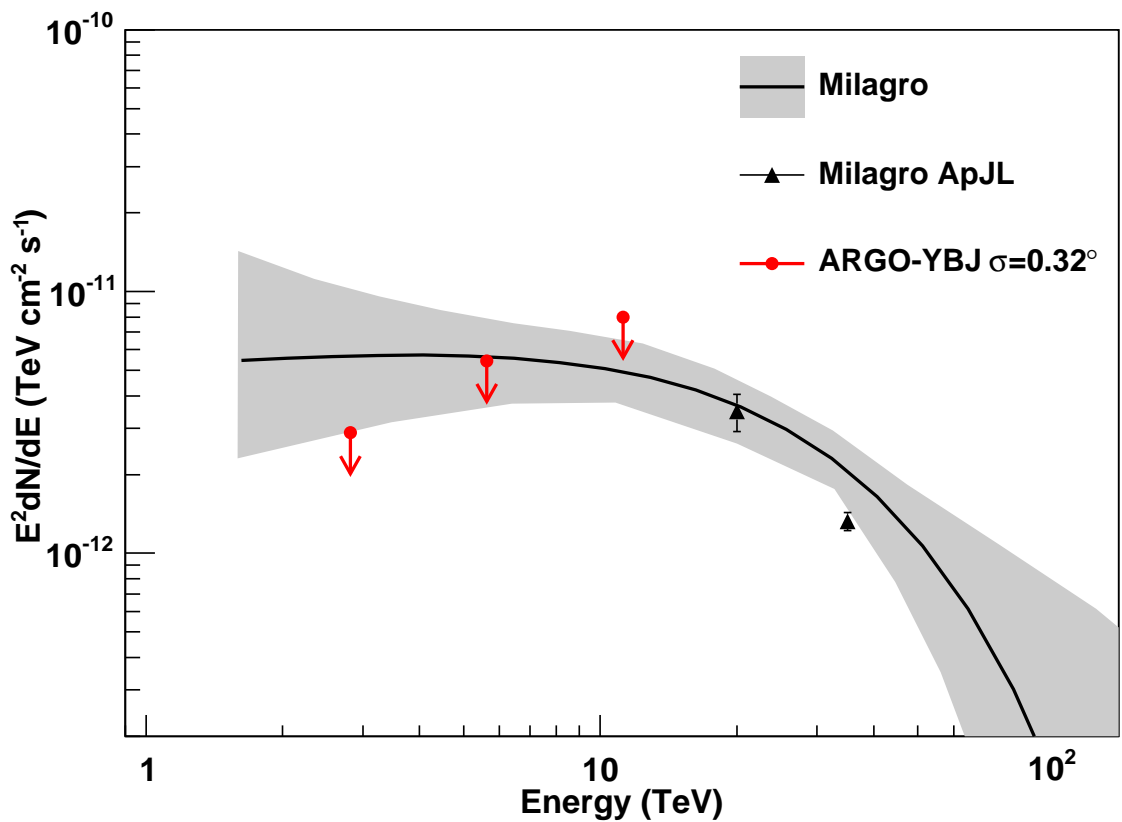}
\caption{Upper limits to the flux from MGRO J2019+37 derived by the ARGO-YBJ experiment adopting the spectrum given in \citep{smith09}. The extension is assumed to be $\sigma=0.32^{\circ}$  as given in \citep{abdo07b}. The solid line and shaded area indicate the differential energy spectrum and the 1 s.d. error region as determined by the Milagro experiment \citep{smith09}. The two triangles give the
previous flux measurements by Milagro at 20 TeV \citep{abdo07a} and
35 TeV \citep{abdo09}. }
\label{fig4}
\end{figure}
\clearpage

\end{document}